\documentclass[showpacs,showkeys,preprintnumbers,amsmath,amssymb]{revtex4}

\usepackage{epsfig}

\def\be{\begin{equation}}

\def\ee{\end{equation}}

\def\lsim{\raise0.3ex\hbox{$<$\kern-0.75em\raise-1.1ex\hbox{$\sim$}}}

\def\gsim{\raise0.3ex\hbox{$>$\kern-0.75em\raise-1.1ex\hbox{$\sim$}}}

\begin{document}
\preprint{BI-TP 2003/10}
\preprint{BNL-NT-03/7}

\title{Heavy Quark Free Energies and Screening in SU(2) Gauge Theory}

\medskip

\author{S. Digal}
\author{S. Fortunato}

\affiliation{Fakult\"at f\"ur Physik, Universit\"at Bielefeld, D-33501 Bielefeld
, Germany}
\author{P. Petreczky}\thanks{Goldhaber Fellow}
\affiliation{Nuclear Theory Group, Department of Physics, Brookhaven National Laboratory,
Upton NY 11973-500}

\begin{abstract}
\noindent
We study the properties of the  free energy of infinitely heavy
quark anti-quark pair in SU(2) gauge theory. 
By means of
lattice Monte Carlo simulations we calculated the free energies
in the singlet, triplet and color averaged channels,
both in the confinement
and in the deconfinement phase. The singlet and triplet
free energies are defined in Coulomb gauge which is equivalent
to their gauge invariant definitions recently introduced 
by Philipsen.
We analyzed the short and the long
distance behavior, making comparisons
with the zero temperature case. The temperature dependence
of the electric screening
mass is carefully investigated. The order
of the deconfining transition is manifest in the
results near $T_c$ and it allows a reliable test
of a recently proposed method to renormalize the Polyakov loop.

\end{abstract}



\maketitle

\vskip0.7cm

\section{Introduction}

The study of the free energy of a static heavy quark-antiquark pair
in a medium of color charges at some temperature $T$ has recently become
a hot topic of statistical $QCD$ \cite{okacz00}-\cite{zantow03}.
The main reason is the fact that such free energy is related to 
the potential between quarks and antiquarks, 
which is of fundamental importance
both for  understanding deconfinement 
and in heavy quark phenomenology at finite temperature \cite{MS,DPS}.
The study of static quark free energies (Polyakov loop correlators)
is also important for constructing effective theories  at
the deconfinement transition \cite{pisarski}

The presence of the medium affects the quark-antiquark 
interaction in a non trivial way. 
Such modifications of interactions is usually studied in terms of the free
energy of a static quark-antiquark pair separated by some distance
$R$. So far most studies concentrated on the behavior of
the static quark-antiquark free energy at large distances,
$R \gg 1/T$ \cite{okacz00,engels87,irback91};  
below the deconfinement
temperature $T_c$ such behavior is characterized 
by a temperature
dependent string tension, above $T_c$ by exponential screening which is 
governed by a
temperature dependent (Debye) screening mass.
However, it turned out that  the free energies of static quark-antiquark 
exhibit a quite complex temperature behavior already
at short distances $R<1/T$ \cite{attig88,engels89,petr02,okacz02}.
Furthermore, the detailed study of the free energy at short distances
allows to define a renormalized order parameter \cite{okacz02}.
As far as the physics of heavy quarkonia is concerned, the detailed structure
of the static quark-antiquark free energy at short distances is even
more important than its large distance behavior. Though the large distance
behavior of the color averaged potential was extensively studied in Refs. 
\cite{okacz00,engels87,irback91} the problem was not completely settled. This
is partly due to finite size effects and very large statistics needed in such
studies.

In this paper we want to study various features of the heavy quark free energy
in $SU(2)$ gauge theory at finite temperature.
The study of
$SU(2)$ lattice gauge theory has at least two advantages:
the simulations are not so time
consuming as in $SU(3)$ and the deconfinement transition 
is second order, which has interesting consequences on
the behavior of the
free energy near the critical temperature $T_c$.
In general, the heavy quark free energy depends on
the color channel one considers; for a complete analysis we
investigated the singlet, the triplet
and the color averaged channels. In particular the study of the
color singlet free energy is of interest because it
is the most relevant quantity as far as the physics of
heavy quarkonia at finite temperature is concerned.
We examined both the short and the long-distance behavior
of the free energies, below and above the deconfinement temperature. 
We devoted a special attention to the issue of screening, in that
we determined the Debye masses at various temperatures.

The rest of the paper is organized as follows. In section II
we define the free energy of a static quark-antiquark pair in
color singlet, color triplet and color averaged channels and
discuss the choice of the simulation parameters. The basic
features of the static quark-antiquark free energies are
also discussed there. In section III we present the numerical
results below $T_c$. Section IV deals with free energies
above deconfinement and determination of the screening masses.
In section V we define the renormalized Polyakov loop for
SU(2) gauge theory following Ref. \cite{okacz02}. Finally
section VI contains our conclusions.

\section{Free Energies in SU(2) Gauge Theory}

On the lattice the free energy of a static quark-antiquark
pair in the gluonic medium is determined by  
correlation functions of temporal Wilson lines $L$,
\begin{equation}
L(\vec{R})=\prod_{\tau=0}^{N_{\tau}-1} U_0(\vec{R},\tau),
\end{equation}
$Tr L$ is also referred to as the Polyakov loop.
Following Refs. \cite{mclerran81,nadkarni86} we introduce
the color singlet and triplet free energy of a static
quark-antiquark pair
\begin{equation}
e^{-F_{1}(R,T)/T+C}=
\frac{1}{2}{\langle\,Tr\,(L(\vec{R})L^{\dagger}(\vec{0}))\rangle},
\label{F1}
\end{equation}
\begin{equation}
e^{-F_{3}(R,T)/T+C}=
\frac{1}{3} {\langle Tr L(\vec{R}) Tr L^{\dagger}(\vec{0})\rangle}-
\frac{1}{6} {\langle Tr L(\vec{R}) L^{\dagger}(\vec{0})\rangle}
\label{F3}
\end{equation}
as well as the color averaged free energy defined by
\begin{equation}
e^{-F_{avg}(R,T)/T+C}=\frac{1}{4}
{\langle\,Tr\,L(\vec{R})Tr\,L^{\dagger}(\vec{0})\rangle}.
\end{equation}
The latter can be written as a thermal average of the free
energies in singlet and triplet channels, hence the name,
\begin{equation}
e^{-F_{avg}(R,T)/T}=\frac{1}{4} e^{-F_1(R,T)/T}+\frac{3}{4} e^{-F_3(R,T)/T}.
\label{av}
\end{equation}
The normalization constant $C$ can be defined in different ways. In
the deconfined phase it is customary to set $C=\ln {|\langle \frac{1}{2} Tr L \rangle|}^2$.
Another possibility is to fix it by normalizing the singlet free energy
to the zero temperature heavy quark potential \cite{okacz02}.

The main problem with the definitions of the singlet and triplet
free energies (\ref{F1})-(\ref{F3}) is that these definitions are
not gauge invariant, as the Wilson line is not a gauge invariant
quantity. The only manifestly gauge invariant quantity is the 
color averaged free energy. This is the reason why singlet and
triplet free energies were not studied in much detail so far.
It was recently showed by Philipsen that gauge invariant definitions 
of the singlet and triplet free energies can be achieved by replacing
the Wilson line in Eqs. (\ref{F1}), (\ref{F3}) by a gauge
invariant Wilson line defined by
\begin{equation}
\tilde L(\vec{R})= \Omega^{\dagger}(\vec{R}) L(\vec{R}) \Omega(\vec{R})
\end{equation}
The $SU(2)$ matrix $\Omega(\vec{R})$ is constructed from eigenvectors of the spatial
covariant Laplacian (see Ref. \cite{ophil02} for further details).
Furthermore it was shown that this definition is equivalent to the
definitions of the singlet and triplet free energies in Coulomb gauge.
Since the determination of eigenvectors of the covariant Laplacian
is computationally very expensive we fix the Coulomb gauge
to calculate the singlet and triplet free energies.

In our numerical investigations we use the standard Wilson action.
In  order to get control over finite size effects, which become
important in the vicinity of $T_c$, we have performed simulations
at several different volumes. 
As we also want  to investigate the short distance behavior of the
free energies simulations were performed for $N_{\tau}=4,6,8$.
To fix the temperature scale we have used the non-perturbative 
beta function of Ref. \cite{engels95}. We will also use 
$T_c/\sqrt{\sigma}=0.69$ \cite{fingberg93}, with $\sigma$ being
the zero temperature string tension. The lattice volumes and the
gauge coupling along with the corresponding 
temperatures used in our simulations are summarized in Table 1.

\begin{table}[htbp]
\begin{center}
\begin{tabular}{c c c c c c c c c c c c}
\hline
\multicolumn{3}{c|}{$N_\tau=4$} &
\multicolumn{3}{c|}{$N_\tau=6$} &
\multicolumn{3}{c|}{$N_\tau=8$} &
\multicolumn{3}{c}{$N_\tau=12$} \\
\hline
$\beta$ & $T/T_c$ & $N_{\sigma}$ &$\beta$ & $T/T_c$ & $N_{\sigma}$ &$\beta$ &
$T/T_c$ & $N_{\sigma}$ 
&$\beta$ & $T/T_c$ & $N_{\sigma}$\\
\hline
2.1962 & 0.70 & 32 & 2.5000 & 1.30 & 32 & 2.4781 & 0.90 & 32 & 2.5000 & 0.60 &
32 \\ 
2.2340 & 0.80 & 16,32 & & & & 2.7385 & 2.00 & 32 & & & \\
2.2681 & 0.90 & 16,32 & & & & 2.8765 & 3.00 & 32 & & & \\
2.2745 & 0.92 & 32,60 & & & & 3.1228 & 6.062 & 32 & & & \\
2.2807 & 0.94 & 32 & & & & 3.2218 & 8.00 & 32 & & & \\
2.2838 & 0.95 & 32,60 & & & & 3.3680 & 12.00 & 32 & & & \\
2.2900 & 0.97 & 32,60 & & & & & & & & & \\
2.2930 & 0.98 & 60 & & & & & & & & & \\
2.2960 & 0.99 & 32,60 & & & & & & & & & \\
2.2975 & 0.995& 32 & & & & & & & & & \\
2.3019 & 1.01 & 32,48,60 & & & & & & & & & \\
2.3077 & 1.03 & 32 & & & & & & & & & \\
2.3134 & 1.05 & 16,32,48 & & & & & & & & & \\
2.3272 & 1.10 & 32 & & & & & & & & & \\
2.3533 & 1.20 & 16,32 & & & & & & & & & \\
2.3776 & 1.30 & 32 & & & & & & & & & \\
2.4215 & 1.50 & 16,32 & & & & & & & & & \\
2.5118 & 2.00 & 16,32 & & & & & & & & & \\
2.6431 & 3.00 & 32 & & & & & & & & & \\
2.8800 & 6.062 & 32 & & & & & & & & & \\
2.9766 & 8.000 & 32 & & & & & & & & & \\
3.0230 & 9.143 &32 & & & & & & & & & \\
3.2190 & 15.87 & 16 & & & & & & & & & \\
\hline
\end{tabular}
\caption{\label{tab0}
Lattice volumes and gauge couplings $\beta=4/g^2$ used in our
simulations. The values of $T/T_c$ were obtained using the non-perturbative 
beta function \cite{engels95}.}
\end{center}
\end{table}

        \begin{figure}[htb]     
\centerline{
        \epsfxsize=9.cm\epsfbox{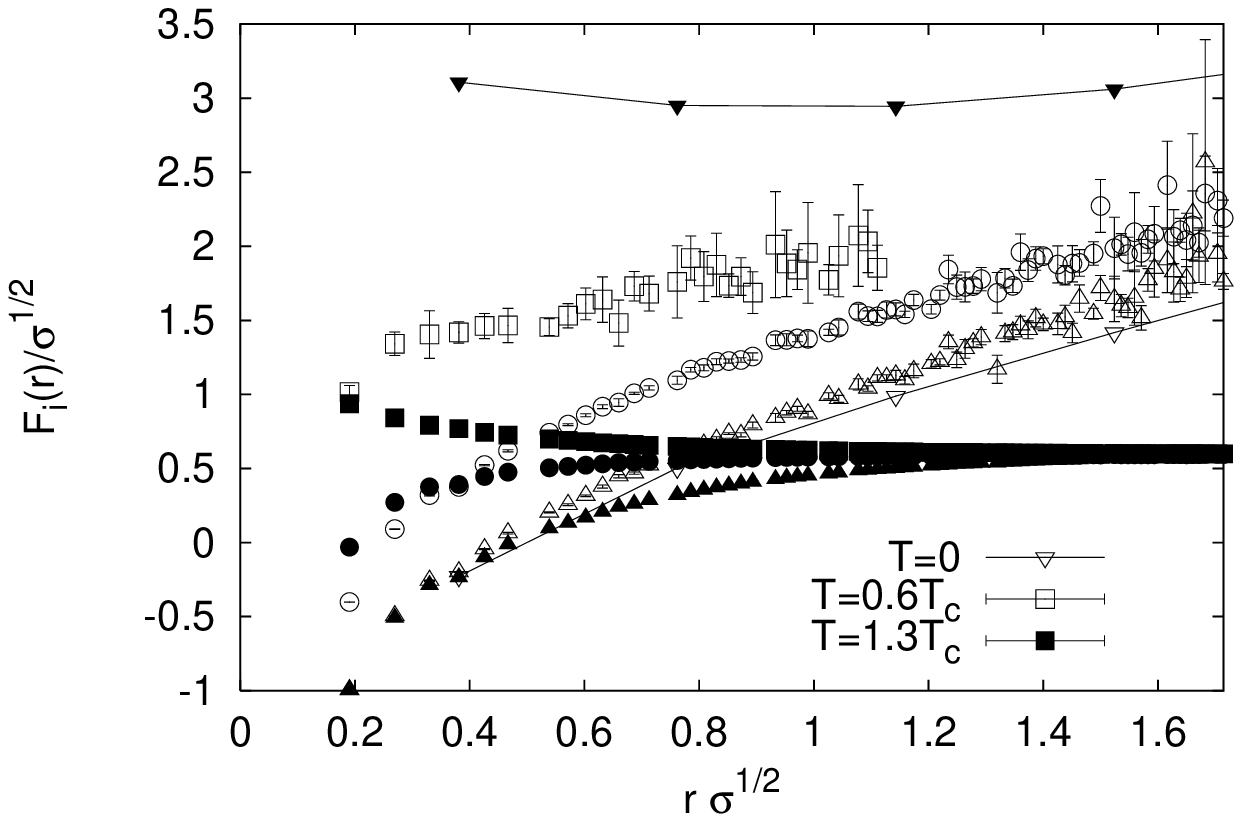}
        \epsfxsize=9.cm\epsfbox{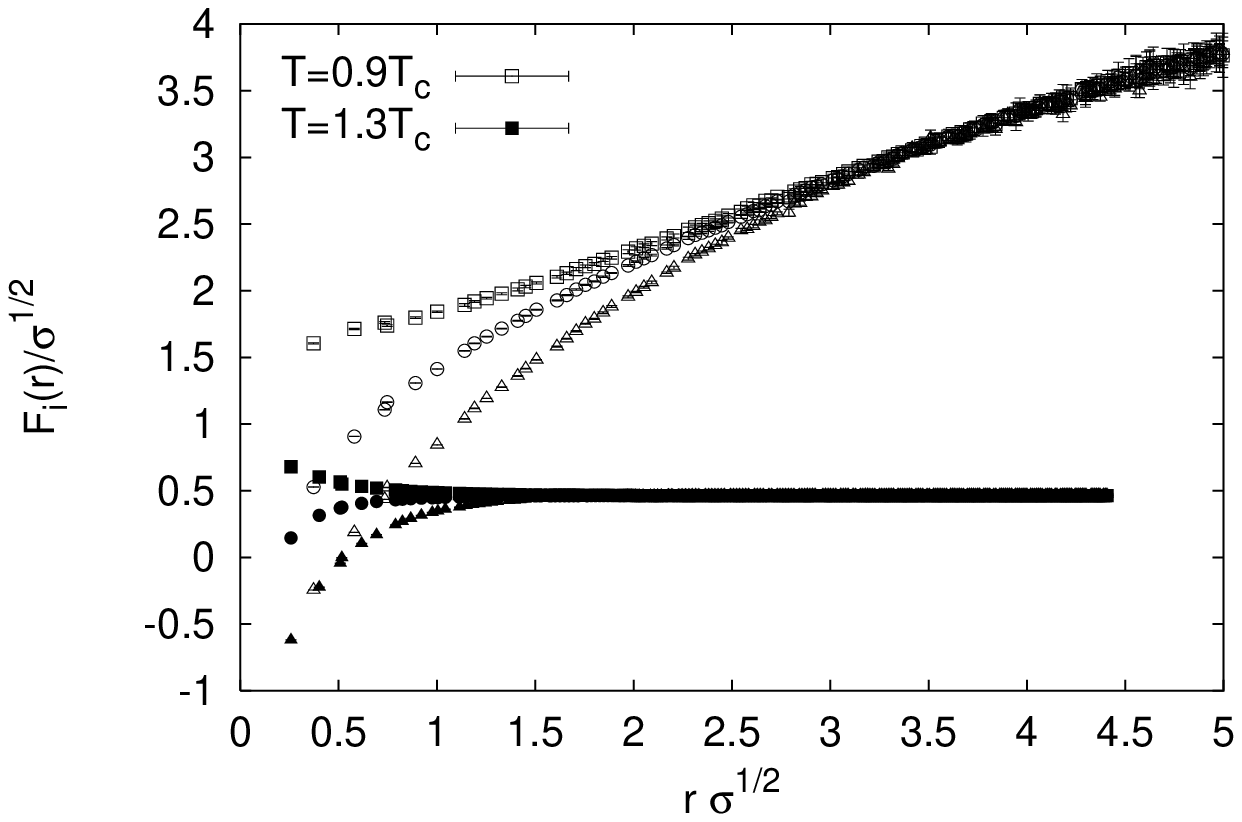}}
        \vspace*{-0.2cm}
        \caption{(Left) Free energies of a static quark-antiquark pair 
          as a function of the distance $r\sqrt{\sigma}$ 
        at $T=0.6T_c$ (open symbols) and $T=1.3T_c$ (filled symbols) corresponding to
         coupling $\beta=2.5$ Also shown there is the $T=0$ potential 
         and its first excitation (open and filled lower triangles correspondingly
         connected by lines) at the same  $\beta$ value. (Right)
         The free energies are here calculated at $N_{\tau}=4$; the temperature
         values are $T=0.9T_c$ and $T=1.3T_c$, respectively. 
         The triangles, squares and 
          circles indicate the singlet, the triplet, and the averaged  
          free energies, respectively.\label{fig1}}
        \end{figure}

Calculations of the zero temperature potentials and 
of the free energy of a static quark-antiquark pair
show violation of rotational symmetry at short distances.
Since we are also interested  in the behavior of
the potential at short distances, we should try to 
remove these lattice artifacts. Following Ref. \cite{necco02}
we replace $F_i(r)$ by $F_i(r_I)$ where 
\begin{equation}
r_I^{-1}=4 \pi \int \frac{d^3 k}{(2 \pi)^3} \exp(i \vec{k} \cdot \vec{r})
\frac{1}{\sum_{i=1,3} \sin^2 (k_i/2)}
\label{rI}.
\end{equation}
In this way we replace the lattice separation by the separation
$r_I$ which corrects for the tree level artifacts in the
Coulomb potential calculated on the lattice. When presenting the
data on the free energy we will always do this replacement unless
stated otherwise.

Let us now present some general features of our
findings.
First we have performed simulations at a fixed lattice spacing, corresponding to the 
gauge coupling $\beta=2.5$ at $N_{\tau}=12$ and $N_{\tau}=6$ corresponding
to temperatures
$T=0.6T_c$ and $T=1.3T_c$, respectively. The results are shown in Fig. \ref{fig1} (left).
At this value of the gauge coupling the ground state static quark-antiquark potential
as well as the first two excited 
potentials at $T=0$ have been calculated \cite{perantonis} and we 
show them in Fig. \ref{fig1}
together with our data.
The singlet free energies
at short distances do not differ from the zero temperature potential
and temperature dependence shows up only at $r \sqrt{\sigma}>0.5$.
On the other hand we find that the triplet
free energy is considerably smaller than the
first excited potential at $T=0$. At small distances it was shown
that the excited potential coincides with the perturbative 
octet (triplet) potential up to a non-perturbative constant \cite{brambilla00}.  
No such constant is expected in our definition of the triplet
free energy which we expect to coincide with the perturbative
one at short distances. At very short distances also the triplet 
free energy is temperature independent.

In most of our calculations we have varied the temperature $T$
 by varying the lattice spacing $a$ (i.e. the gauge coupling $\beta=4/g^2$) 
for fixed temporal extent
$N_{\tau}$. As both the free energy and the $T=0$ potential
contain a lattice spacing dependent additive constant $C$ 
(c.f. Eqs. \ref{F1}-\ref{av}), some normalization prescription
should be introduced in order to compare the free energies 
calculated at different temperatures. To do so we assume
the following form for the zero temperature potential
\begin{equation}
V(r)/\sqrt{\sigma}=-\frac{0.238}{r \sqrt{\sigma}}+ r \sqrt{\sigma}
+\frac{0.0031}{(r \sqrt{\sigma})^2}
\label{vT=0}
\end{equation}
This form was obtained in Ref. \cite{booth} by fitting
the lattice data on the $T=0$ potential for $r \sqrt{\sigma}>0.063$ 
at $\beta=2.85$ apart from
the constant which we have omitted. The violation of rotational
invariance on the lattice was taken into account in the fit procedure. Thus
equation (\ref{vT=0}) defines our convention for the continuum zero temperature
potential. In some cases we need the $T=0$ potential at distances $r \sqrt{\sigma}<0.063$.
In this case we use the 3-loop perturbative potential calculated in $qq$ scheme 
\cite{necco01} normalized to smoothly match the form defined by Eq. (\ref{vT=0}) at 
$r \sqrt{\sigma}=0.07$. We have also checked that the difference between the 3-loop
and 2-loop results is negligible for our purposes. In what follows the normalization
constant $C$ will be chosen such (unless stated otherwise) 
that the singlet free energy matches the $T=0$ 
potential at the shortest distance ($r/a=1$).
In Fig. \ref{fig1} (right) we show our results on static free energies calculated for
$N_{\tau}=4$ and using this normalization convention.
The free energies in the deconfined phase reach the same
value at large distances (see Fig. \ref{fig1}). 
This is to be expected as at very
large distances due to screening 
the free energy of color charges should be independent of
their relative color orientation. What is more
interesting is that at large distances the gap between the triplet
and singlet free energy vanishes as well in the confinement phase though it
is non-zero for the $T=0$ case. This issue will be discussed more 
in detail in the next section.

\section{Results in the Confinement Phase}

\begin{figure}[htb]     
\centerline{
\epsfxsize=9.cm\epsfbox{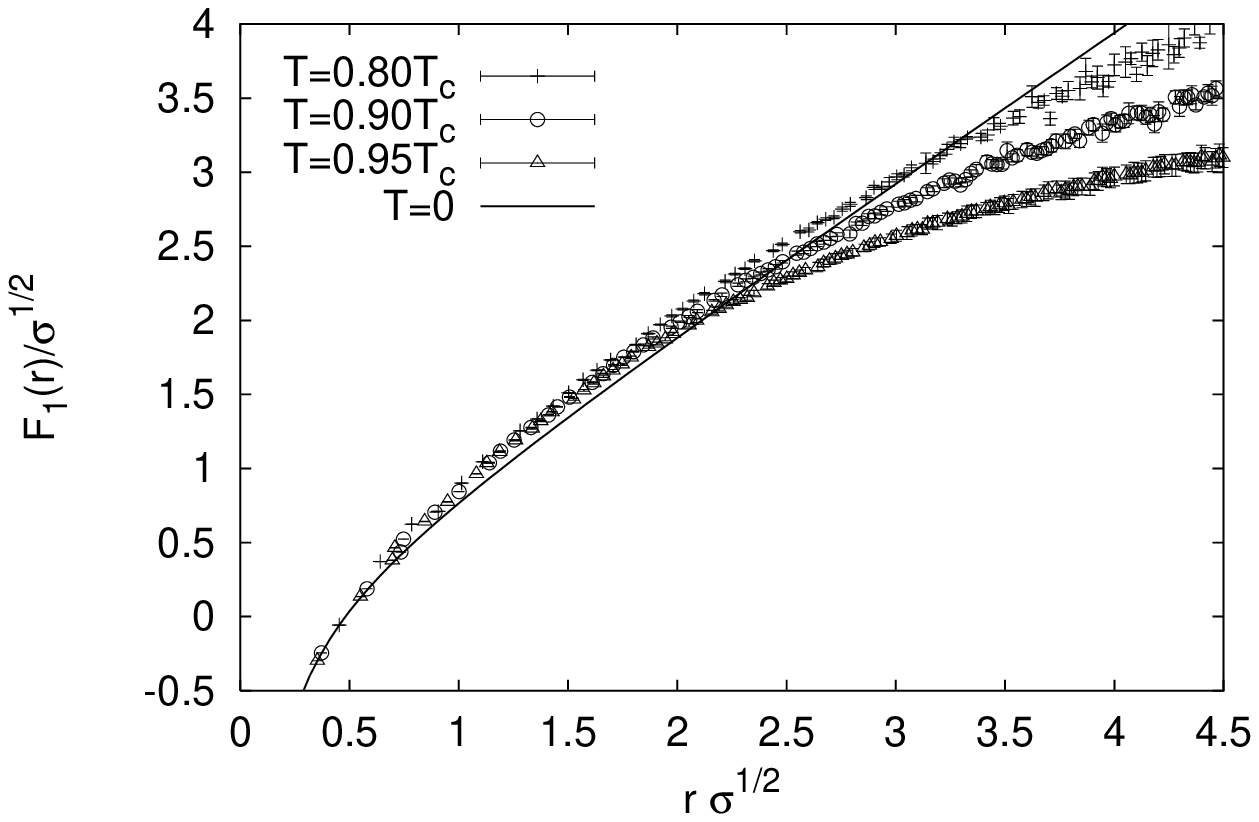}   
\epsfxsize=9.cm\epsfbox{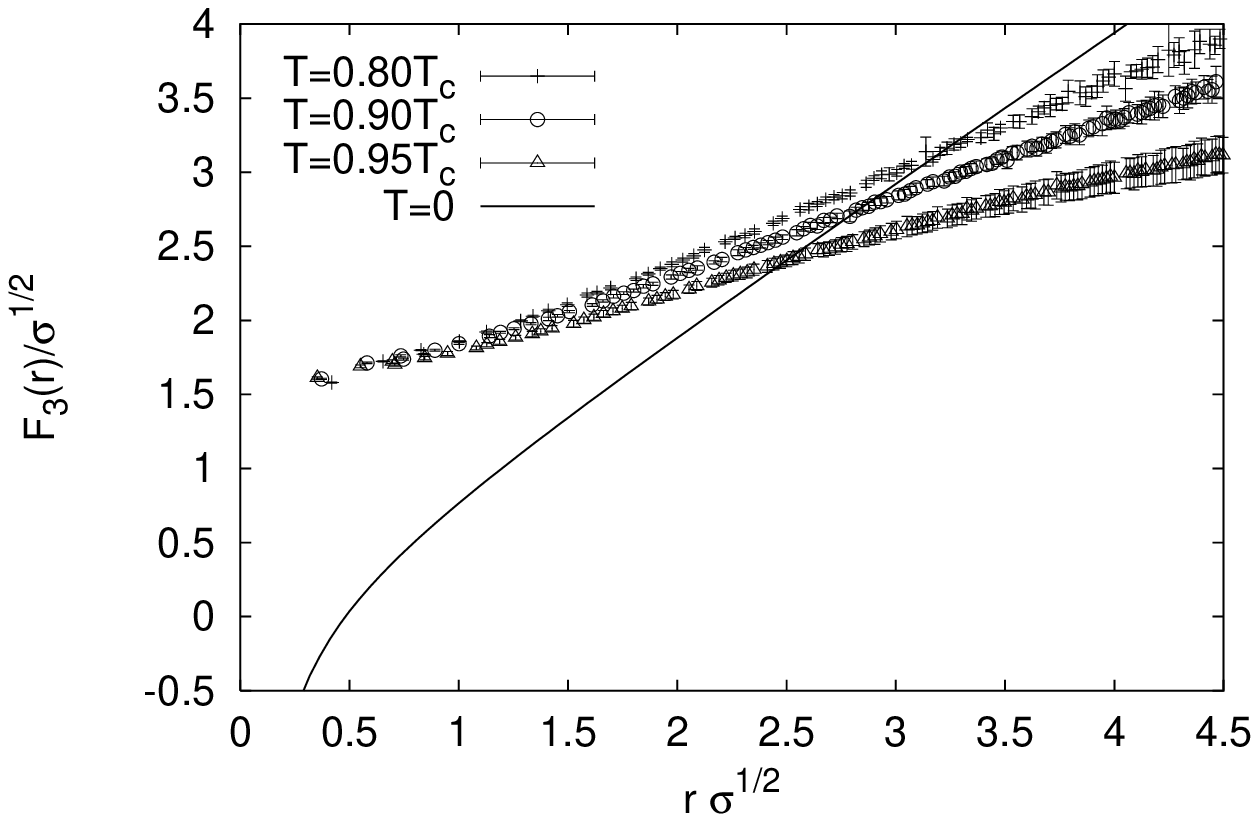}   
}
\vspace*{-0.2cm}
\caption{The color singlet (left) and triplet (right)
free energies
at various temperatures in the confinement phase
calculated for $N_{\tau}=4$. The normalization constant 
$C$ was chosen such that the singlet free energy matches
the zero temperature potential (solid line) at the shortest distance
(see text).
}
\label{fig2}
\end{figure}
\begin{figure}[htb]     

\centerline{
\epsfxsize=9.cm\epsfbox{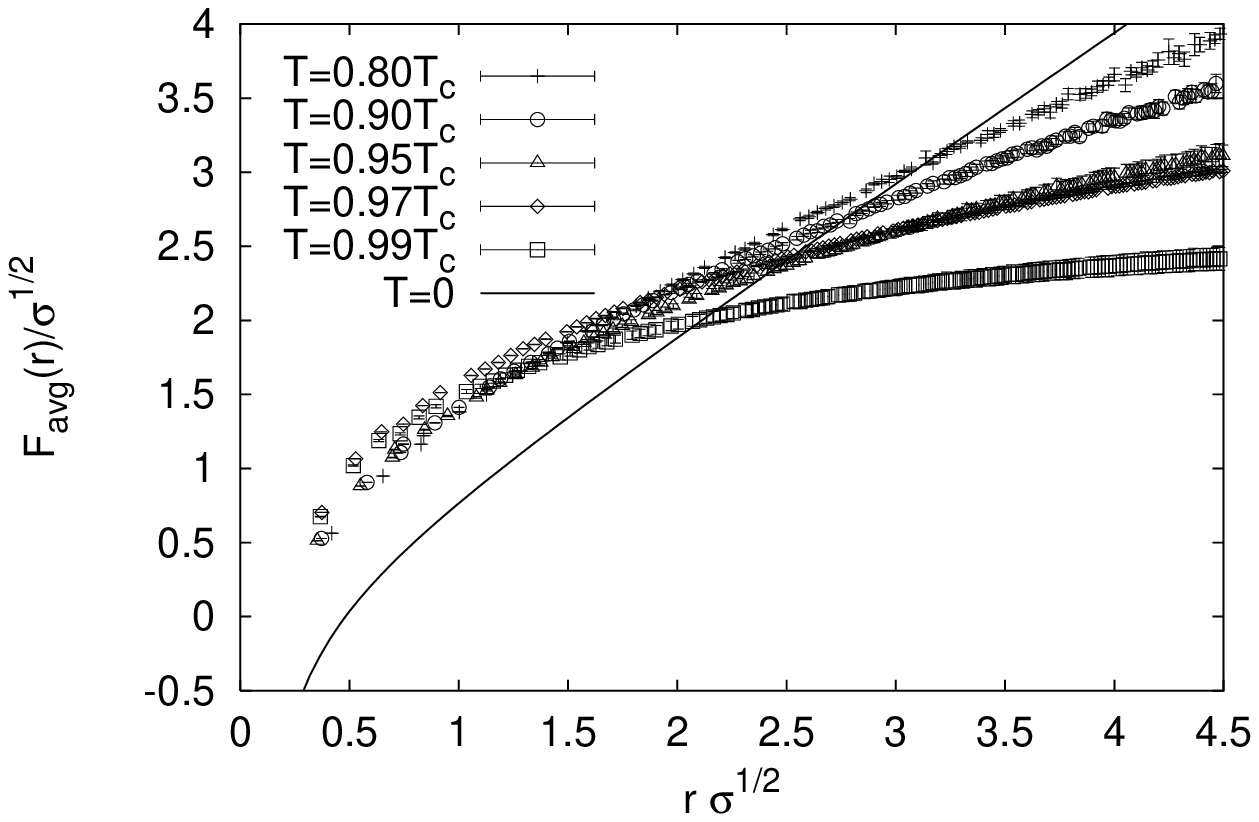}   
\epsfxsize=9.cm\epsfbox{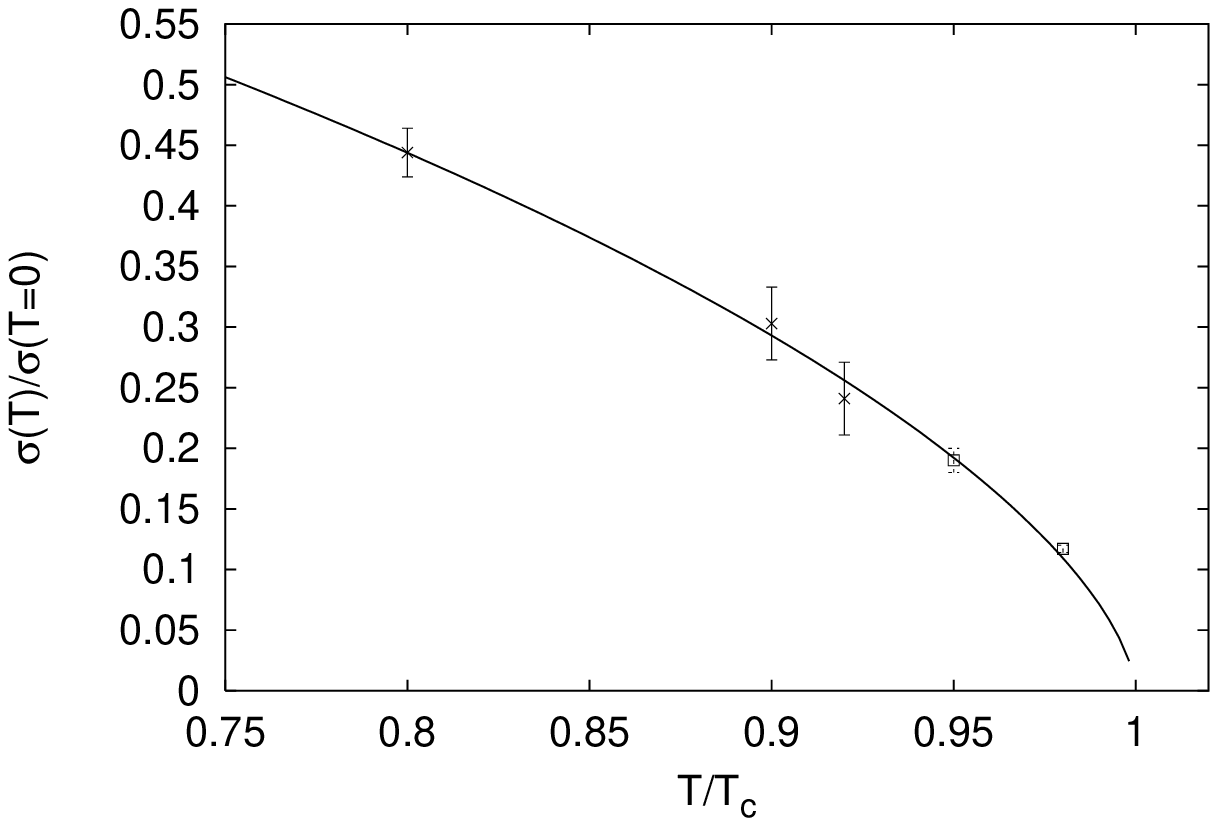}  
} 
\vspace*{-0.2cm}
\caption{(Left) Color averaged free 
energies at various temperatures in the confinement phase. (right)
String tension as a function of $T/T_c$, crosses correspond to $32^3 \times 4$
lattice, open squares to $60^3 \times 4$ lattice. 
\label{avgstring}}
\end{figure}

In this section we are going to present our numerical results
in the confinement phase. In Fig. \ref{fig2} we show
the singlet and triplet free energies with the normalization
described in the previous section. One can see that the temperature
dependence of the singlet free energy is only visible at
distances $r \sqrt{\sigma}>2$. The triplet free energy is 
also temperature independent at small distances;
thermal effects become visible at $r \sqrt{\sigma}>1.4$.
We also notice that there is a slight enhancement of the singlet
free energy over the $T=0$ potential in the interval
$1 < r \sqrt{\sigma} < 2$. A similar enhancement was observed also
in the case of $2+1$ dimensional $SU(2)$ gauge theory at 
finite temperature \cite{ophil02} as well as in preliminary $SU(3)$ 
calculations \cite{zantow03}.
As the color averaged free energy is a thermal average of singlet and
triplet free energies, it would have a non-trivial temperature
dependence even if the latter were temperature independent. This
temperature dependence is larger the smaller the gap between
the triplet and singlet contributions. In general one can say
that the temperature dependence of the color averaged free energy
at short and intermediate distances 
is mostly determined by the value and the temperature 
dependence of the color triplet contribution. If this continues
to be true in full QCD, then some of the conclusions of 
Ref. \cite{DPS} (where the color averaged free energy was related to
the meson masses) should be revised.

At large distances the color singlet, triplet and averaged free 
energy  reach a common value which can be parametrized
by a form
\begin{equation}
F_i(r,T)|_{r T \gg 1}=\sigma(T) r + A(T) \ln r T +B(T)
\label{fitbelow}
\end{equation}
In Fig. \ref{avgstring} we show the color averaged free energy as
well as the string tension obtained from it using the fit
to Eq. \ref{fitbelow}. Because the deconfinement transition is
of second order, the inverse correlation length (string tension)
vanishes at $T_c$. As a result we have large finite size
effects close to $T_c$. Indeed we found that on the $32^3{\times}4$ lattice
the string tension vanishes around $0.97T_c$.
For this reason we performed simulations on the larger
$60^3{\times}4$ lattice close to $T_c$. 
We have seen that 
the values of $\sigma(T)$ we have obtained from our calculations 
on $32^3 \times 4$ lattice agree with the
results from the $60^3{\times}4$ lattice 
up to $0.92T_c$. Above such temperature we adopt the results 
from the larger lattice. The final situation is illustrated
in Fig. \ref{avgstring} (right), where the squares indicate the results 
from $60^3{\times}4$, the crosses the ones from $32^3{\times}4$.
It is well known that $\sigma(T)$ vanishes near $T_c$ according
to the power law $\sigma(T)\,\propto\,(T_c-T)^\nu$, where $\nu=0.63$ 
is the 3D Ising exponent for the correlation length. Therefore we tried to
fit our data point by using the 
ansatz $\sigma(T)=a(T_c-T)^{\nu}[1+b(T_c-T)^{1/2}]$, with $\nu=0.63$.
The best fit curve is shown in our plot and it reproduces very well our data.
The values of $\sigma(T)$ found by us are considerably smaller than those
obtained in Ref. \cite{engels87}. This is probably due to the fact that
the lattice volumes used in Ref. \cite{engels87} were considerably smaller
than ours.

\section{Results in the Deconfinement Phase: Screening}

\begin{figure}[htb]     
\begin{center}
\centerline{
\epsfxsize=8.2cm\epsfbox{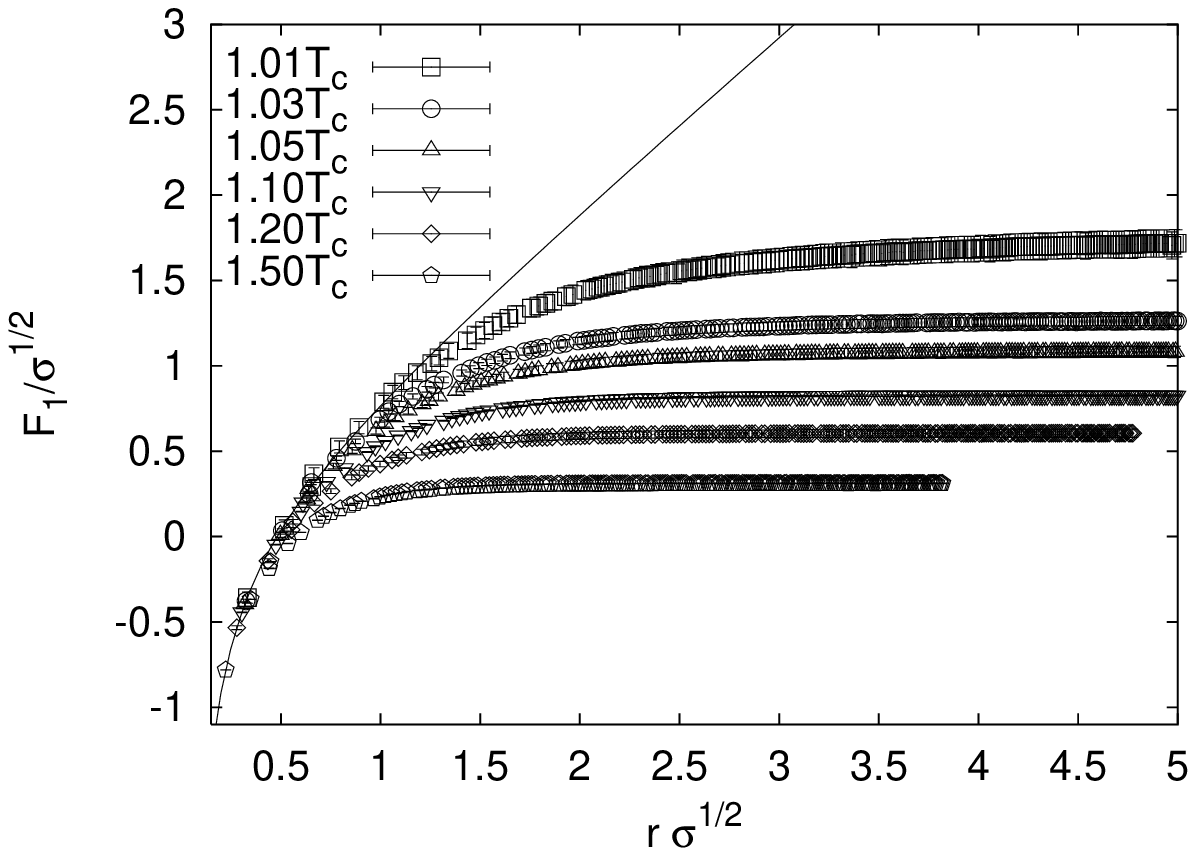},
\epsfxsize=8.2cm \epsfbox{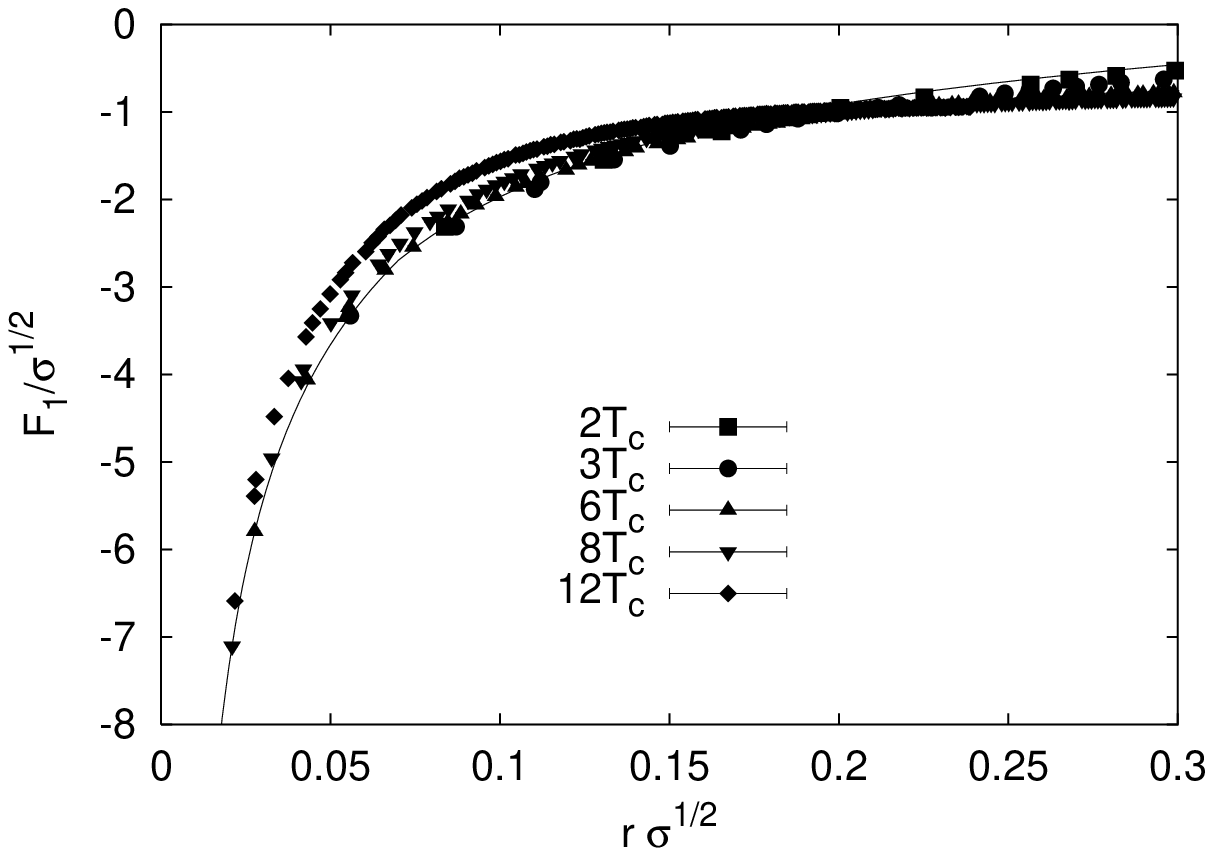}
}   
\vspace*{-0.2cm}
\caption{The color singlet
free energy
at various temperatures in the deconfinement phase
calculated for $N_{\tau}=4$ (left) and $N_{\tau}=8$ (right).
The solid line is the $T=0$ potential.}
\label{f1above_ph}
\end{center}
\end{figure}

We start our discussion of the numerical results in the
deconfined phase with Fig. \ref{f1above_ph}, where we show
the singlet free energy in units of $\sqrt{\sigma}$ normalized to
the zero temperature potential at the shortest distance. As one can see the singlet free
energy saturates at large distances, while at short enough distances
it is temperature independent and coincides
with the zero temperature potential. 
The distance in physical units, at which the temperature dependence
enters, of course strongly depends on the value of the temperature, the higher
the temperature the shorter is the distance where effects of the medium become 
visible. Also the distance where the singlet free energy saturates strongly depends
on the temperature, it is getting larger as we approach $T_c$. Close to $T_c$
screening enters only at distances $r \sqrt{\sigma} >1$. This feature of the
free energy is reflected in the temperature dependence of the screening masses
which will be discussed below.
Another interesting feature
of the singlet free energy is that the value of the plateau of
the free energy decreases with the temperature. Such a
behavior of the singlet free energy was observed for $SU(3)$ gauge theory
in Ref. \cite{okacz02}, where it was also argued that the reason for this is
the presence of the entropy contribution. 

\begin{figure}[htb]     
\begin{center}
\epsfxsize=8.2cm\epsfbox{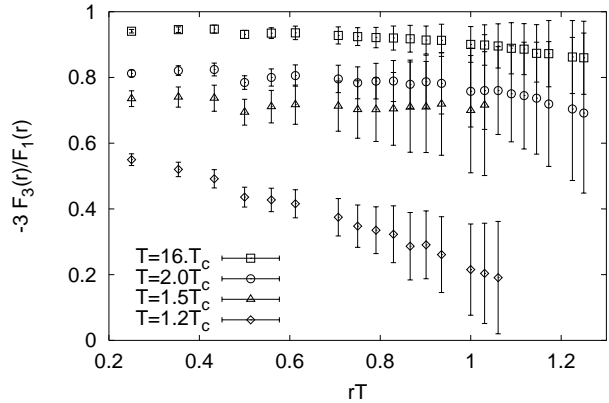}  
\vspace*{-0.2cm}
\caption{
Free energy triplet/singlet ratio at various temperatures.\label{f3suf1}}
\end{center}
\end{figure}

For the further discussion of the results in the deconfined phase,
especially for making comparisons with perturbation theory,
it is more convenient and
in fact customary (cf. \cite{okacz00,engels89})
to choose the renormalization
constant $C$ in Eqs. (\ref{F1}), (\ref{F3}) to be
$C=\ln |\langle \frac{1}{2} Tr L \rangle|^2$.
Furthermore the large distance behavior of the free
energies should be discussed separately from their
short distance behavior, where one would expect
perturbation theory to work. In general it is expected
that perturbation theory breaks down at distances
$r > 1/g^2 T$ \cite{linde80}, with $g$ being
the gauge coupling constant. As for the physically
interesting temperature range one always has $g \sim 1$,
perturbation theory may be applicable at distances
$rT <1$. One of the predictions of perturbation theory
is that $-3 F_3(r,T)/F_1(r,T) \simeq 1$, for any $r$.
In Fig. \ref{f3suf1} we show this ratio for different
temperatures. This ratio appears indeed to be constant for
$T>1.5T_c$ but always smaller than $1$, even for the 
highest temperature we considered ($16T_c$). 
Similar results were found in Landau gauge in Refs. \cite{attig88, heller98}.

\begin{figure}[htb]     
\begin{center}
\centerline{
\epsfxsize=9cm\epsfbox{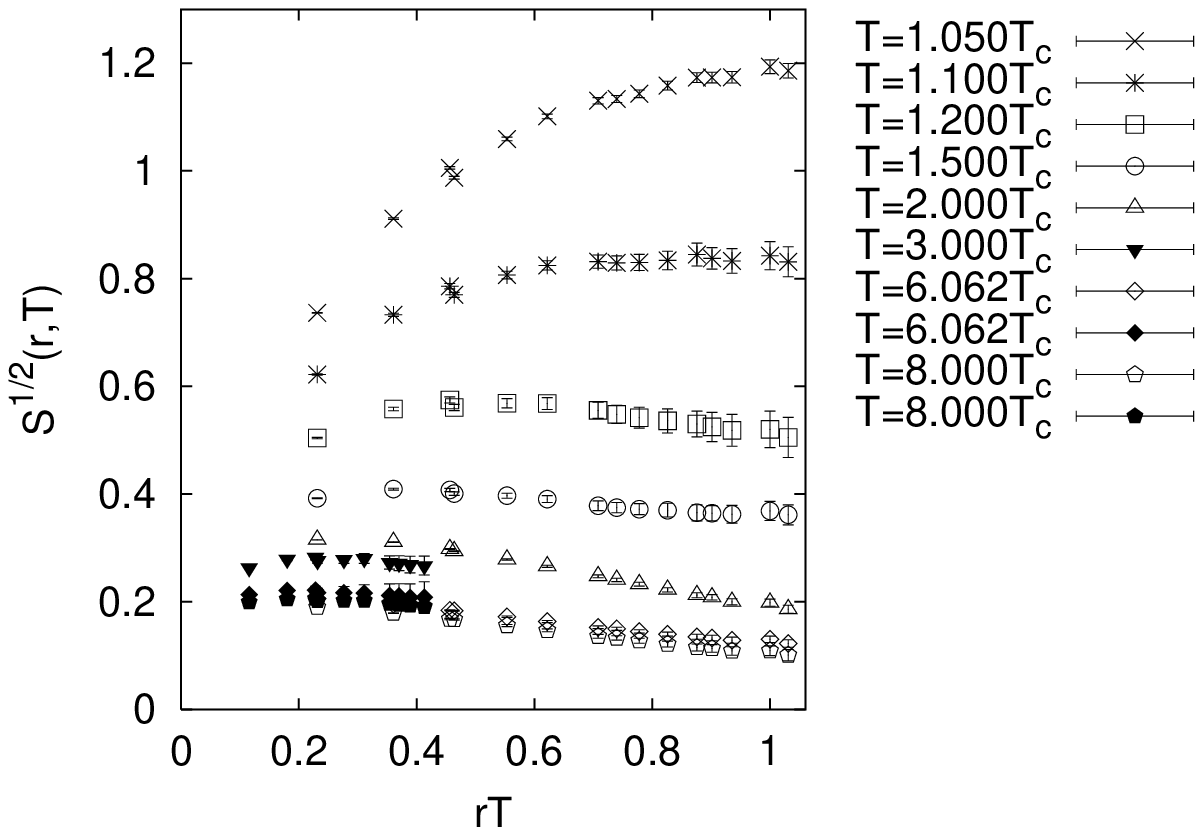}
\epsfxsize=9cm\epsfbox{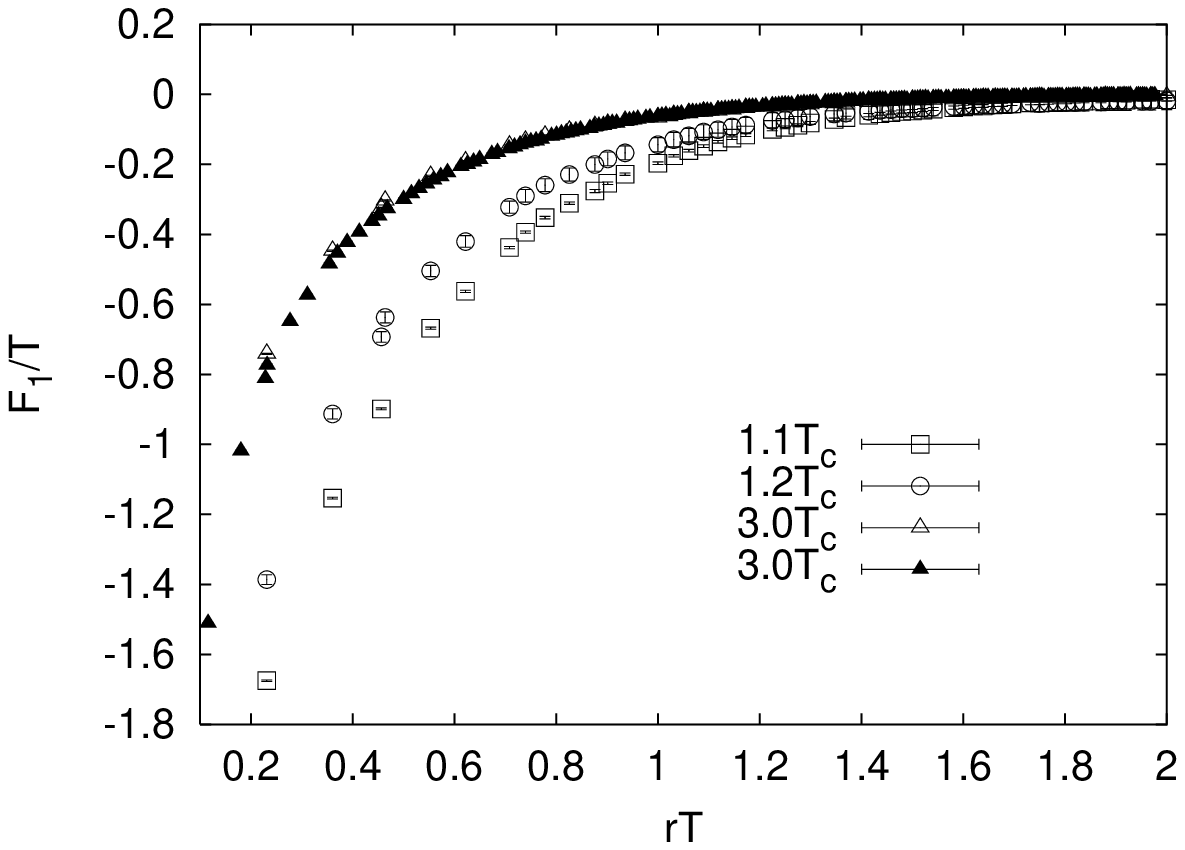}
}
\vspace*{-0.2cm}
\caption{
The square root of the screening function (left) and the singlet free
energy in units of $T$ (right) at
different temperatures. The open symbols refer to the results for $N_\tau=4$,
the full symbols to the results for $N_\tau=8$.}
\label{scf}
\end{center}
\end{figure}

High temperature perturbation theory predicts that
the color averaged free energy has the form \cite{mclerran81,nadkarni86a}
\begin{equation}
\frac{F_{avg}(r,T)}{T}=-\frac{3}{32}\frac{g^4}{(4 \pi r T)^2} e^{- 2 m_{D0} r}
\label{pform}
\end{equation} 
at distances $r>1/T$, with $m_{D0}$ being the leading order
Debye mass $m_{D0}=\sqrt{2/3} g T$. At distances $r < 1/T$ 
the simple form (\ref{pform}) is no longer valid, though
the $1/r^2$-like behavior is still expected due to cancellation
between singlet and triplet contributions \cite{mclerran81}.
Therefore we define the so-called screening function $S(r,T)$
by the formula
\begin{equation}
\frac{F_{avg}(r,T)}{T}=-\frac{3}{32}\frac{1}{(rT)^2} S(r,T)
\end{equation}
In Fig. \ref{scf} we show the numerical results for 
the square root screening function. As one can
see from the figure, at high temperatures ($T>2T_c$)
the screening function shows a mild $r$-dependence, which implies
that the color averaged free energy behaves like $1/r^2$.
The screening function $S(r,T)$ decreases with with increasing
thenperature which one would expect if $S(r,T) \sim g^4(T)$.
At temperatures closer to $T_c$ the screening function 
decreases at small distances. Obviously, this behavior
has nothing to do with screening and signals the breakdown
of the high temperature expansion. As we approach $T_c$
$F_{1,3}/T$ is no longer small at small distances as shown in Fig. 6 
and therefore the  exponentials in
Eq. (\ref{av}) cannot be expanded. As a result of this,
the cancellation between the singlet and triplet
free energies no longer holds; moreover, since the
singlet free energy is negative and the triplet one is positive
(when working with normalization convention $C=\ln |\langle \frac{1}{2} Tr L \rangle|^2$)
the color averaged free energy is dominated by the singlet contribution
and behaves like $1/r$ at small distances.
We also note that the free energies calculated for different $N_{\tau}$
(different lattice spacing) at $3T_c$ agree reasonably well. Similar 
agreement between the results calculated for different values of $N_{\tau}$
was observed for other temperatures too.

Let us now discuss the large distance
behavior of the free energies and the determination of the screening masses.
We will restrict ourselves to the discussion of the color singlet
and color averaged free energy as the color triplet free energy becomes very noisy at
large distances and the present statistics does not allow to study it
in detail.

Contrary to earlier studies \cite{okacz00,engels87,engels89},
where the screening 
masses were obtained
using uncorrelated fit and the short and large distance behaviors
of the free energy were not separate, here we use the correlated fit
procedure of Ref. \cite{mckerr95}. 
From Fig. 6 it is clear that in the region $r T<1$ the color averaged
free energy can be well described by almost unscreened $1/r^2$-like
behavior. Therefore this region should not be considered for the
determination of the screening masses. In our procedure
the fit interval was chosen so that 
the fit yields a reasonable $\chi^2/(\# D.o.F)$. Thus, unlike 
in the uncorrelated fit used in Ref. \cite{okacz00}, there is
no dependence of the screening masses on the fit interval. 
We determine the screening mass in the color singlet channel
by fitting the data with a screened Coulomb (Yukawa-like) 
ansatz. In leading order perturbation theory, in fact, 
the most important contribution to the singlet free energy
is given by the exchange of a single gluon. In the case 
of the color average free energy we used a more general fit ansatz
\footnote{In fact in order to do the correlated fit we consider
the connected correlators of Wilson lines, i.e. we subtract 
$|\langle Tr L\rangle|^2$ from the correlator. As far as we are interested
only in the large distance behavior this is equivalent to fitting the
free energies. Only in the thermodynamic limit the connected correlator
vanishes at very large distance. Therefore we allow for a constant $B$
in our fit ansatz. It turns out, however, that this constant is
compatible with zero within present statistical accuracy.}:
\begin{equation}
\frac{F_{avg}}{T}\,=\,\frac{A}{R^d}\exp(-\mu{R})+B.
\label{fitanzatz}
\end{equation}
As possible values for the exponent $d$ we took $d=1,2$.
The resulting values of the screening masses
that we extracted are presented in Tables \ref{tab1} and \ref{tab2}
for the singlet and the averaged channels, respectively.
For determination of the screening masses we mostly use $32^3 \times 4$
lattice.

We found that for the color averaged free energy the fits with 
$d=1$ and $d=2$ are both good for all temperatures, 
except near $T_c$, where 
we got a reasonable $\chi^2/(\# D.o.F)$ only for $d=1$.
From our fit analysis it then comes out
that we cannot choose between the two cases, which gives 
a systematic error of about $30\%$ on the screening masses. 
In order to eliminate this ambiguity, we have also calculated
the plane-plane correlator, given by the formula:
\begin{equation}
C_{PL}(x_3)={\langle\,Tr\,L(x_3)Tr\,L^{\dagger}(0)\rangle}-|\langle Tr L \rangle |^2
\end{equation}
where $L(x_3)\,\equiv\,\sum_{x_1,x_2}L(x_1,x_2,x_3)$. 
If the color averaged free energy has the form (\ref{fitanzatz}) with $d=1$, then
$C_{PL}(x_3)$ should fall off 
with the distance as a simple exponential, which allows a 
direct determination of the screening mass. The results we found are 
reported for comparison in Table \ref{tab2}. We see that the values of the 
masses extracted from the plane-plane correlator agree in each case 
with the masses obtained from the point-point correlator when $d=1$.
Furthermore, we have analyzed the effective masses extracted from
$C_{PL}(x_3)$. They reach a plateau already at $x_3 T \sim 1$, which makes
the presence of power-like prefactors in $C_{PL}(x_3)$ at large distances
very unlikely and thus implying that very likely $d \simeq 1$.
At high temperature dimensional reduction arguments
suggest that $d=1$, as the large distance behavior of any static
correlators is governed by exchange of a bound state of the effective
three dimensional theory \cite{dimred}.

For $T=1.01T_c$ 
the results refer to the $48^3{\times}4$ lattice instead of
$32^3{\times}4$.
In this case, in fact,  
the mass changes appreciably when one goes to the larger 
$48^3{\times}4$ lattice. This is probably due to the large correlation
length close to $T_c$. Further simulations on a $60^3{\times}4$ 
lead to the same value of the mass we found on the $48^3{\times}4$, which 
is then reliable as infinite volume limit at this temperature. 

        \begin{figure}[htb]     
        \centerline{
          \epsfxsize=9cm\epsfbox{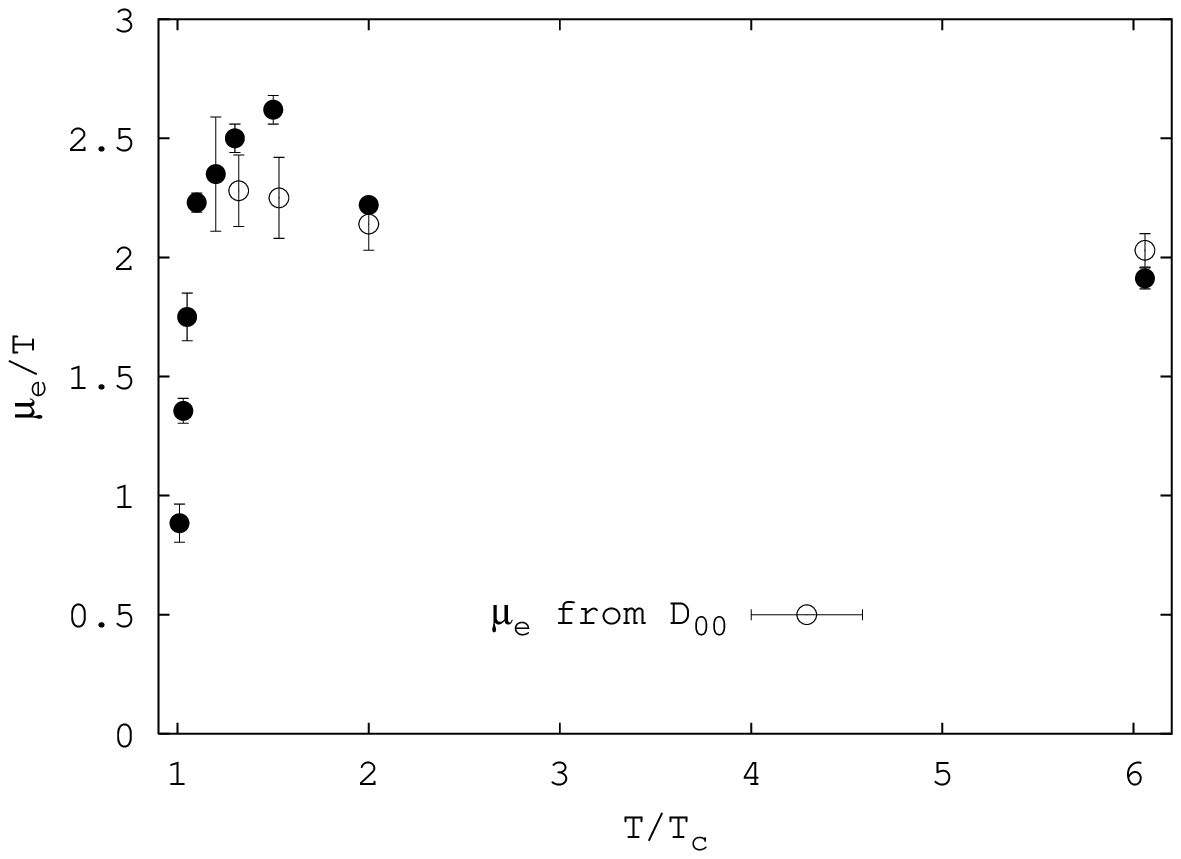}
          \epsfxsize=9cm\epsfbox{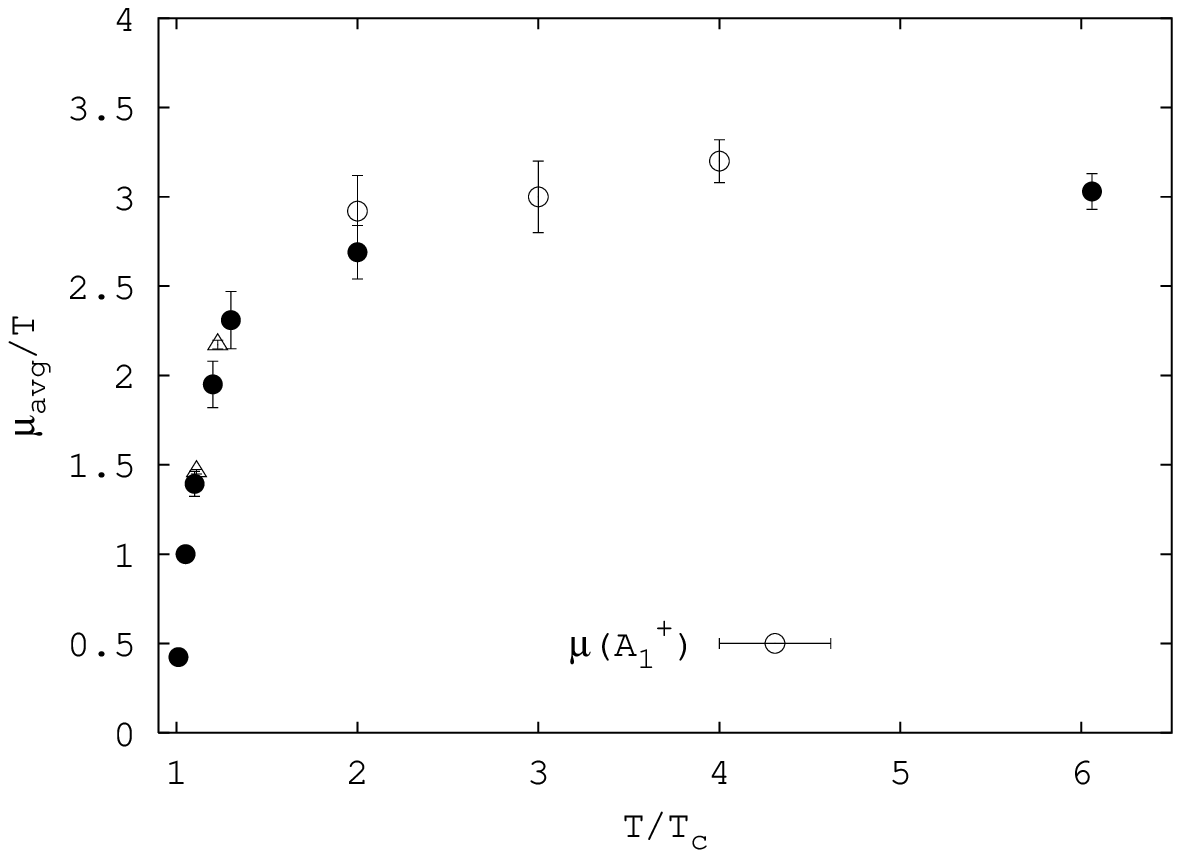}
        }   
        \vspace*{-0.2cm}
        \caption{Screening masses in units of temperature vs.
        $T/T_c$ extracted from the color singlet free energy
        (left) and from the color averaged free energy (right).
        In the left plot we have also shown the screening mass extracted
        from the static electric propagator \cite{heller98,rank}.
        In the right plot we give the lowest $A_1^{+}$
        screening mass from \cite{datta98} as well as the screening masses
        of plane-plane correlators of Polyakov loops from \cite{fiore03}
        (open triangles).}
       \label{mass}
        \end{figure}

In Fig. \ref{mass} we show the screening masses extracted from
the singlet free energy
as a function of the temperature.
There we also show the values of the screening masses obtained from
the electric gluon propagator in Landau gauge 
\cite{heller98,rank} at $T >1.2T_c$.
As one can see they are compatible with the singlet masses we have found.
This is to be expected in perturbation theory. But as $g \sim 1$ in the
temperature interval considered and the screening mass governs the
large distance behavior of the free energy, this agreement is quite
non-trivial.
In Fig. \ref{mass} the color averaged screening masses are shown as well;
we compared them with the lowest $A_1^{+}$ scalar screening mass 
(spatial glueball mass) obtained in Ref. \cite{datta98}. 
Dimensional reduction arguments \cite{dimred}
suggest that these masses should agree at high temperature and the figure 
seems to indicate that this is indeed the case.
Very recently the plane-plane correlators of Polyakov loops were 
studied in Ref. \cite{fiore03} for $T=1.1105T_c$ and $T=1.227 T_c$
(we use the non-perturbative beta function \cite{engels95} to convert the gauge
coupling of Ref. \cite{fiore03} to temperature).
In Fig. \ref{mass} we show as well the corresponding screening masses (open
triangles). 
Moreover we note that the color averaged screening masses
obtained by us using fits with $d=2$ agree with findings of
Refs. \cite{irback91,reisz93} where the same exponent was used.
The color averaged screening mass should go to zero when $T{\rightarrow}T_c$
because it is just the inverse of the Polyakov loop correlation
length, which diverges at $T_c$. On the other hand there is no a priori 
argument for which the singlet mass should vanish at the threshold.
Fig. \ref{mass} indicates that both masses become very small near 
$T_c$.

\section{The renormalized Polyakov loop}

\begin{figure}[htb]     
        \centerline{\epsfxsize=9cm\epsfbox{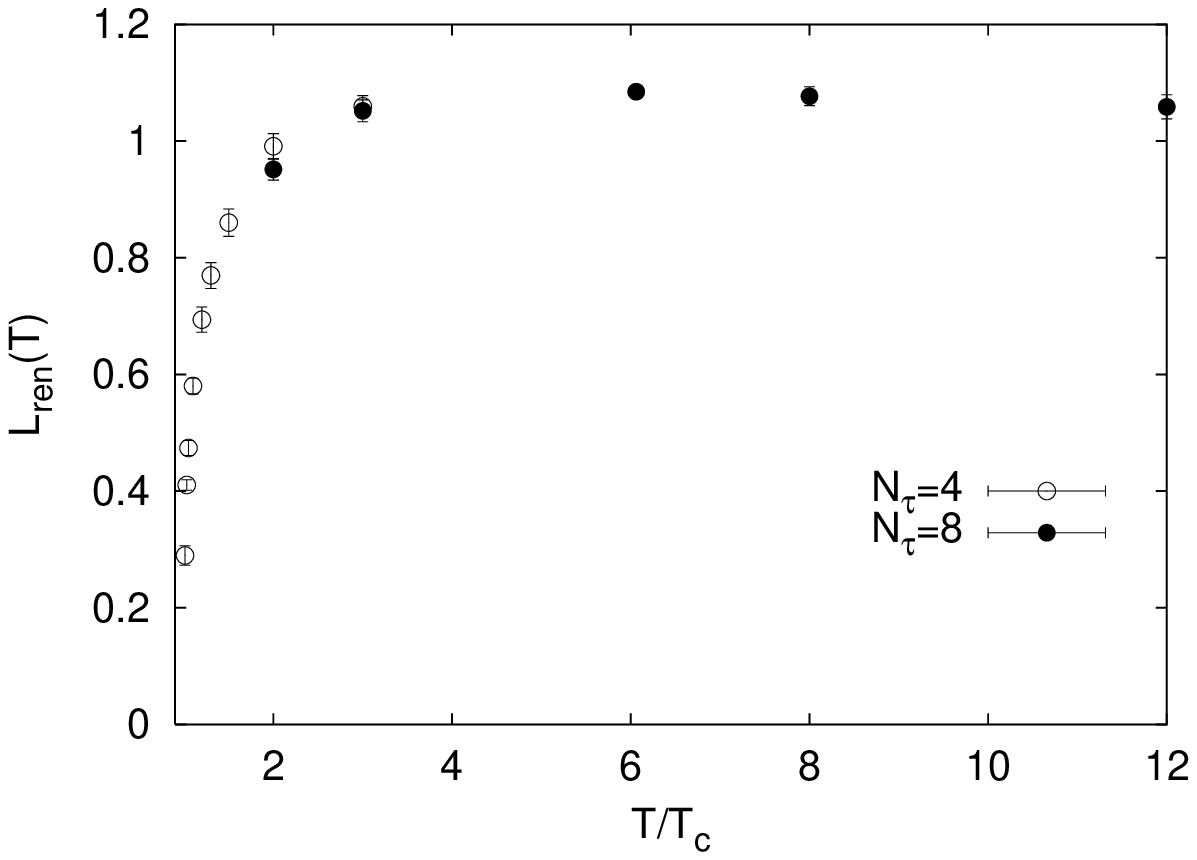}
        \epsfxsize=9cm\epsfbox{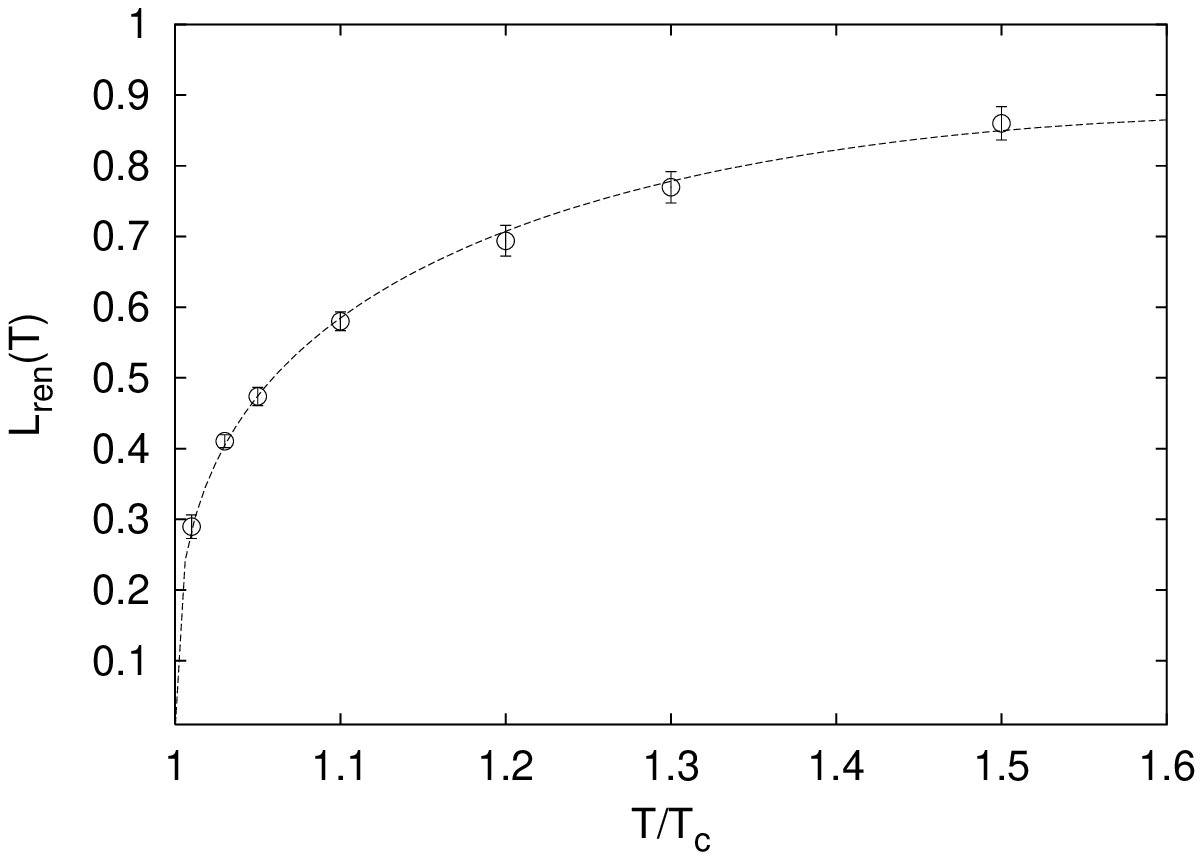} }   
        \vspace*{-0.2cm}
        \caption{Renormalized Polyakov loop as a function of the 
        temperature in the entire temperature interval (left) and
        for $T \le 1.5T_c$ together with the fit (right).
        The error bars indicate the combined statistical and systematic errors.
        \label{lrenorm}}
        \end{figure}

The last issue we would like  to address is the renormalization of 
the Polyakov loop.
It is known that if one takes the continuum limit 
at fixed temperature, the expectation value of the Polyakov 
loop vanishes, so that the usual definition does not
really provide a physical order parameter for deconfinement.
As we have already said at the beginning, the free energies  
on the lattice are always defined up to some renormalization
constant. According to a recent work \cite{okacz02}, a suitable
choice of such renormalization constant can lead to a 
new definition of the Polyakov loop. If the constant is chosen
such that the singlet free energy matches the zero temperature
heavy quark potential at short distances, one can define
a "renormalized" Polyakov loop $L_{ren}$ through the formula:
\begin{equation}
L_{ren}=\exp(-F_\infty(T)/2T),
\label{lren}
\end{equation}
where $F_\infty(T)$ is the asymptotic value of the 
singlet free energy at the temperature $T$
(in fact, it does not depend on relative color orientation of the quark-antiquark pair,
see above).
We remind that we have renormalized the singlet free energy exactly in this
way, so that, in our case, $F_\infty(T)$ is nothing but
the height of the plateau of the curves in Fig. 4.
Practically we took $F_\infty(T)=F_1(N_{\sigma}/2,T)$, which is the
value of the free energy at the largest distance allowed on the lattice.
In Fig. \ref{lrenorm} we plot the renormalized Polyakov loop
as function of the temperatures.
When normalizing the singlet free energy to the $T=0$ potential
at the shortest distance ($r/a=1$) we implicitly assume that there is no  
temperature dependence at this distance. This may not  always be  the 
case. Therefore we also calculate $F_{\infty}(T)$ and the corresponding
$L_{ren}$ by normalizing the singlet free energy to the $T=0$
potential as well at $r/a=\sqrt{2}$. The difference in $F_{\infty}(T)$ 
( $L_{ren}$) arising from these two normalizations give us an estimate of 
possible systematic errors. When quoting the error on the renormalized
Polyakov loop we always add quadratically the systematic and 
the statistical errors.
At variance with $SU(3)$, we now have a second order phase transition
and a well defined scaling behavior of $L_{ren}$ at criticality.
In order to check the scaling we have fitted the data on $L_{ren}$
using  the standard ansatz in the interval $T_c < T \le 1.5T_c$

\begin{equation}
L_{ren}(T)=c(T-T_c)^{\beta}[1+b(T-T_c)^{\omega}],
\end{equation}

with the exponents $\beta$ and $\omega$ fixed to their
$SU(2)$ values $\beta=0.3265$ and $\omega=1$.
The fit curve is shown in Fig. \ref{lrenorm} (dashed line) and reproduces
quite well the pattern of the data points. Considering only temperatures
$T<1.1T_c$ it is possible to fit the data with $b=0$ and approximately
same value of $c$.

\section{Conclusions}

In conclusion we have studied the quark-antiquark 
free energies in $SU(2)$ gauge theory below and above
the deconfinement temperature. We have found that the 
temperature dependence of the singlet free energy is much
weaker than the temperature dependence of the color 
averaged free energy. Most of the temperature dependence
of the color averaged free energy is due to the presence of
the color triplet contribution and its temperature dependence.
If this will hold for real QCD it may have important consequences
in the heavy quarkonia phenomenology at finite temperature.
At large distances the free energy in color singlet and triplet
channel converge to a common value.

Above $T_c$ the color singlet free energy can be understood
in terms of  propagation of a non-perturbatively screened gluon.
Through the detailed analysis of the point-point and
plane-plane correlators as well as comparisons with
other determinations of the screening masses we have established
that
the color averaged free energy is described by Yukawa law
at large distances whereas
at shorter distances it exhibits 
a more complex behavior. At low temperatures and short distances the color 
averaged free energy is dominated by the singlet contribution,
while at higher temperatures it has $1/r^2 T$ behavior and
its temperature dependence is qualitatively the same as
predicted by perturbation theory. 
Finally we have shown that the renormalized Polyakov loop
defined in \cite{okacz02} has the correct scaling behavior near the  
critical temperature.

As an outlook 
we note that we have studied the free energy of 
static charges in fundamental representation. It will be interesting 
to investigate the free energy of static charges in other representations.
Some work in this direction was done in \cite{redlich88}.

\begin{table}[htbp]
\begin{center}
\begin{tabular}{|c|c|c|}
\hline
\multicolumn{3}{|c|}{Colour Singlet Correlators} \\
\hline\hline
\multicolumn{1}{|c}{$\beta$} &
\multicolumn{1}{|c}{$T/T_c$} &
\multicolumn{1}{|c|}{$\mu_{e}(T)/T$} \\
\hline
2.3019 & 1.01  &  0.884(80) \\
\hline
2.3077 & 1.03  &  1.356(52) \\ 
\hline
2.3134 & 1.05 & 1.75(10) \\
\hline
2.3272 & 1.10 & 2.23(4) \\
\hline
2.3533 & 1.20 & 2.35(24) \\
\hline
2.3776 & 1.30 & 2.50(6) \\
\hline
2.4215 & 1.50 & 2.62(6) \\
\hline
2.5118 & 2.00 & 2.37(9) \\
\hline
2.8800 & 6.062 & 1.912(44)\\
\hline
3.0230 & 9.143 & 1.812(36) \\
\hline
\end{tabular}
\caption{Screening masses extracted from color singlet
free energy. \label{tab1}}
\end{center}
\end{table}

\begin{table}[htbp]
\begin{center}
\begin{tabular}{|c|c|c|c|c|}
\hline
\multicolumn{5}{|c|}{Colour Averaged Correlators} \\
\hline\hline
& & \multicolumn{3}{c|}{$\mu_{avg}(T)/T$, extracted from} \\
\multicolumn{1}{|c}{$\beta$} &
\multicolumn{1}{|c}{$T/T_c$} &
\multicolumn{1}{|c}{Point-C, d=1} &
\multicolumn{1}{|c|}{Point-C, d=2} &
\multicolumn{1}{|c|}{Plane-C} \\
\hline
2.3019 & 1.01  &  0.424(20) & $<0$ & 0.468(28)\\
\hline
2.3134 & 1.05 & 1.000(28) & 0.544(20)& 1.024(16)\\
\hline
2.3533 & 1.20 & 1.95(13) & 1.19(22)& 1.93(6)\\
\hline
2.3776 & 1.30 & 2.31(16) & 1.332(88)& 2.296(64)\\
\hline
2.5112 & 2.00 & 2.69(15) & 2.06(12) & 2.89(18)\\
\hline
2.8800 & 6.062 & 3.03(10) & 2.32(10) &\\
\hline
3.0230 & 9.143 & 3.04(20) & 2.04(56) & \\
\hline
\end{tabular}
\caption{Screening masses extracted from color averaged free
energy and from plane-plane correlators of Polyakov loops
(see text).
\label{tab2}}
\end{center}
\end{table}

\bigskip

\begin{acknowledgments}

It is a pleasure to thank J. Engels, O. Kaczmarek and F. Karsch
for helpful discussions. We are grateful to A. Cucchieri, who provided us
the gauge fixing routine.
We would also like to thank the TMR network ERBFMRX-CT-970122 and 
the DFG Forschergruppe fFOR 339/2-1 or financial support. This work
was partly supported by the U.S. Department of energy under 
Contract DE-AC02-98CH10886.

\end{acknowledgments}

\end{document}